\documentclass[USenglish,twocolumn]{article}

\usepackage[utf8]{inputenc}
\usepackage[big]{dgruyter}
\usepackage{txfonts}
\usepackage{poster}
\usepackage{ulem}

\renewcommand{\ni}{\noindent}
\newcommand{\mc}[3]{\multicolumn{#1}{#2}{#3}}

\begin{document}

  \articletype{Research Article{\hfill}Open Access}

\author*[1]{Dana Kovaleva}
\author[2]{Anatoly Piskunov}
\author[3]{Nina Kharchenko}
\author[4]{Ralf-Dieter~Scholz}

  \title{\huge A preliminary comparison of photometric (MWSC) and trigonometric (TGAS) distances of open cluster stars}

  \runningtitle{Photometric and trigonometric distances of open cluster stars}


  \begin{abstract}
{ Our goal was to compare the open cluster photometric distance scale of the global survey of star clusters in the Milky Way (MWSC) with the distances derived from trigonometric parallaxes from the Gaia DR1/TGAS catalogue and to investigate to which degree and extent both scales agree. We compared the parallax-based and photometry-based distances of 5743 cluster stars selected as members of 1118 clusters based on their kinematic and photometric MWSC membership probabilities. We found good overall agreement between trigonometric and photometric distances of open cluster stars. The residuals between them were small and unbiased up to $\log (d,[pc]) \approx 2.8$. If we considered only the most populated clusters and used cluster distances obtained from the mean trigonometric parallax of their MWSC members, the good agreement of the distance scales continued up to $\log (d,[pc]) \approx 3.3$.}
  \end{abstract}


  \journalname{Open Astronomy}
\DOI{DOI}
  \startpage{1}
  \received{..}
  \revised{..}
  \accepted{..}

  \journalyear{2017}
  \journalvolume{1}

\maketitle


{ \let\thempfn\relax
\footnotetext{\hspace{-1ex}{\Authfont\small
\textbf{Corresponding Author: Dana Kovaleva:}} {\Affilfont Institute of Astronomy, Russian Academy of Sciences, 48 Pyatnitskaya Str., 109017 Moscow, Russia; Email: dana@inasan.ru}}
}

{ \let\thempfn\relax
\footnotetext{\hspace{-1ex}{\Authfont\small \textbf{Anatoly Piskunov:     }} {\Affilfont Institute of Astronomy, Russian Academy of Sciences, 48 Pyatnitskaya Str., 109017 Moscow, Russia; Zentrum f\"ur Astronomie der Universit\"at Heidelberg, Landessternwarte, K\"{o}nigstuhl 12, 69117 Heidelberg, Germany}}
\footnotetext{\hspace{-1ex}{\Authfont\small \textbf{Nina Kharchenko:  }} {\Affilfont  Main Astronomical Observatory, 27 Academica Zabolotnogo Str., 03143 Kiev, Ukraine; Zentrum f\"ur Astronomie der Universit\"at Heidelberg, Landessternwarte, K\"{o}nigstuhl 12, 69117 Heidelberg, Germany}}

\footnotetext{\hspace{-1ex}{\Authfont\small \textbf{Ralf-Dieter~Scholz:      }} {\Affilfont Leibniz-Institut f\"ur Astrophysik Potsdam, An der Sternwarte 16, 14482 Potsdam, Germany}}

}

\section{Introduction}

  The global survey Milky Way Star Clusters \cite[{MWSC;}][]{2012A&A...543A.156K} is a comprehensive study of Galactic star clusters based on a combination of all-sky catalogues 2MASS \cite{cat2MASS} and PPMXL \cite{2010AJ....139.2440R}. Within this project, the kinematic-photometric membership probabilities of stars in each cluster region and basic cluster parameters (including solar-centric distances) were determined for 2859 known in the literature \cite{2013A&A...558A..53K} and 202 newly-discovered \cite{2014A&A...568A..51S, 2015A&A...581A..39Sc} open clusters. These clusters are widespread over the Galaxy (with solarcentric distances up to $15$~kpc and the mode of $2.4$~kpc) and are of great interest for structure and evolution studies of the Galactic disk population. Previous parallax-based cluster distances \cite{rob99, 2001AstL...27..386L} derived from Hipparcos parallaxes \cite{1997ESASP1200.....E}  were given only for local objects and can not be used for evaluating MWSC distances due to their insufficient extent and/or accuracy.  The distances of MWSC clusters are based on the isochrone fitting technique and only indirectly (via evolutionary stellar models) related to the present cosmic distance scale.

  Gaia \cite{2016A&A...595A...1G}, the Hipparcos successor, has all chances to improve in a few years the situation providing high precision data on trigonometric parallaxes of Galactic stars at distances up to 3 kpc from the Sun. Already with the Tycho-Gaia Astrometric Solution (TGAS) included in the first data release, Gaia DR1 \cite{2015A&A...574A.115M, 2016A&A...595A...1G, 2017A&A...601A..19G, 2017A&A...599A..50A}, parallax accuracies of the order of 0.3-0.7 mas were achieved for about 2 mln. stars. These allow us to compare the photometric distances of cluster stars directly with independent trigonometric measurements at distances up to the typical distance of MWSC clusters.

  With this preliminary study (final results are published by Kovaleva et al. \cite{2017AApL}), we provide a first comparison of the open cluster distance scale using photometric cluster distances from the MWSC survey and individual MWSC member distances derived from their TGAS trigonometric parallaxes. In Sect.~\ref{sec:data}, we describe the data used in this study. In Sect.~\ref{sec:CDB}, we perform the comparison and discuss the derived results of the study, and in Sect.~\ref{sec:concl}, we draw basic conclusions following from the comparison.

\section{The data}
\label{sec:data}

The input data for our comparison of distances to open cluster stars came from two independent sources: the MWSC and Gaia TGAS catalogues. The most probable members of MWSC star clusters were cross-identified with the TGAS catalogue. The mean photometric MWSC distances, derived from cluster Colour-Magnitude Diagrams (CMDs), were assigned to each member of a given cluster and then compared with their individual trigonometric distances derived from their TGAS parallaxes. Hereafter, we call the MWSC distances $d_{MWSC}$, derived as a rule from CMD fitting, photometric distances, and those obtained from TGAS parallaxes $d_{TGAS}$ trigonometric distances.

\subsection{The MWSC: data description, limitations, distance scale, membership}

\begin{table}[b]
\caption[]{Summary on data used for comparison.}\label{tab:Data}
\centering
\begin{tabular}{@{}cccc@{}}     
\hline
\hline
\noalign{\smallskip}
Membership&\mc{2}{c}{Number of stars} \\
\cline{2-4}
\noalign{\smallskip}
& MWSC &\mc{2}{c}{TGAS} \\
&       & All & Reliable\\
\noalign{\smallskip}
\hline
\noalign{\smallskip}
$P>60\%$ &785837 &5743  &4990\\
$P>90\%$ &147236 &1058  &931 \\
\hline
\end{tabular}
\end{table}

The MWSC survey provides a comprehensive sample of star clusters of our Galaxy with a variety of well-determined parameters based on accurate photometric and uniform kinematic stellar data gathered from the all-sky catalogues 2MASS and PPMXL. The full sample contains 3208 objects: 3061 open and 147 globular clusters. In this study, we concentrated on Galactic disk objects (open clusters and associations) and refer to them as MWSC clusters. Within 2 kpc, corresponding to the distance one can reach with TGAS data, there were about 1500 MWSC clusters suitable for this study.

For our analysis, we selected stars with membership probabilities $P>60\%$. The probabilities were extracted from the MWSC catalogue of stars in cluster areas \cite{2013A&A...558A..53K,2014A&A...568A..51S,2015A&A...581A..39Sc}, where they were calculated for every star in the vicinity of the centre of a surveyed cluster based on photometric and kinematic criteria as described in \cite{2012A&A...543A.156K}. They were used for the determination of basic cluster parameters such as the apparent size, mean proper motion, and distance. Hereafter we refer to these selected stars as {\it probable} members, whereas we call those with even higher membership probabilities $P>90\%$ {\it definite} cluster members. The respective numbers of used cluster members are given in Table~\ref{tab:Data}.

For all cluster members we assigned the respective photometric distance of the cluster they belong to. Note that the expected variations of individual stellar distances within a cluster are relatively small. If one accepts that the typical cluster size is equal to the average tidal radius of MWSC clusters of about 10 pc \cite[see][]{2013A&A...558A..53K}, then the error would be about 10\% at the smallest available MWSC distances of the order of 100 pc and less than 0.5\% at about 2 kpc, corresponding to the average distance of MWSC clusters. This error lies well within the expected errors of TGAS parallaxes of 0.3-0.7 mas \cite{2015A&A...574A.115M}, which is of the order of 100\% at a distance of 2 kpc. The comparison of MWSC data with independent cluster distances from the literature \cite{2013A&A...558A..53K}  yielded a typical error of MWSC distances of 11\%.

\subsection{Gaia TGAS: cross-identification, dataset characteristics}
\label{sec:TGAS}

   \begin{figure}
   \centering
   \includegraphics[width=8cm]{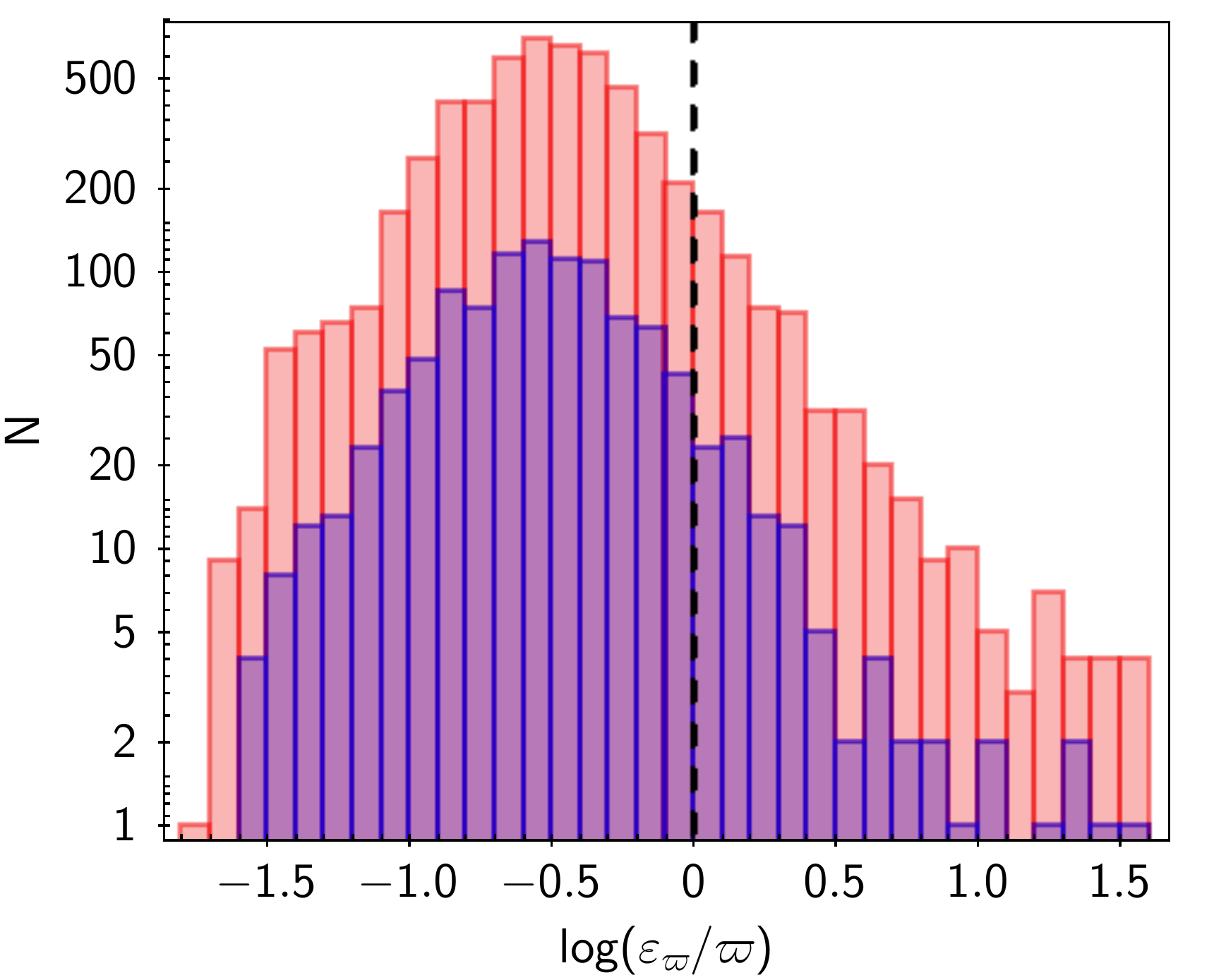}
      \caption{Distribution of  the relative errors of TGAS parallaxes (only for $\varpi>0$) for probable (light red histogram) and definite cluster members (violet histogram). The black vertical dashed line marks relative errors of $\varepsilon_{\varpi}/\varpi =1$. Stars that we considered to have reliable parallaxes (see Sect.~\ref{sec:TGAS}) are located left of this line.
              }
         \label{fig:ParRel}
   \end{figure}

Gaia DR1 is the most precise all-sky astrometric survey since Hipparcos providing TGAS parallaxes for 20 times more stars than known before and with unprecedented precision. However, according to Arenou et al., 2017 \cite{2017A&A...599A..50A}, a global negative parallax zero point (about $-0.04$~mas)  is  consistently  found  with  independent  estimation  methods. These authors recommend to treat Gaia DR1 astrometric data accurately, taking into account a systematic error of about $0.3$ mas in parallaxes. In spite of these limitations, Gaia DR1 provides a new basis for studies of Galactic structure, until it will be superseded by the coming Gaia DR2.

 Our cross-identification of MWSC cluster members with TGAS stars was based on the Sky Algorithm of TopCat \cite{2005ASPC..347...29T}. In total, 5743 probable members of 1118 MWSC clusters were found to have TGAS parallaxes. The first ten of the most populated clusters contain more than 50 probable members with TGAS parallaxes each. For definite members, the numbers are 1058 stars and 481 clusters, respectively. Among them, negative TGAS parallaxes ($\varpi_{TGAS} \leq 0$) were found for 199 probable, and 32 definite cluster members, respectively.

 Figure~\ref{fig:ParRel} shows the distributions of relative parallax errors for probable and definite cluster members. Both distributions behave similarly, showing maxima at about $\log \varepsilon_{\varpi}/\varpi \approx -0.5$ and positive tails corresponding to up to two orders of magnitude larger errors, whereas on the other side the histogram extends to only one order of magnitude smaller errors. The similarity of the distributions implies that both samples are similar to each other with respect to their parallax quality. Therefore, we discuss hereafter mainly the probable cluster members.

 Numbers of stars and parallaxes used in this study are shown in Table~\ref{tab:Data}. All stars with $\varpi>0$ and $\varepsilon_{\varpi}/\varpi < 1$  were considered by us as stars with {\it reliable} parallaxes.

  \begin{figure}
   \centering
   \includegraphics[width=6cm]{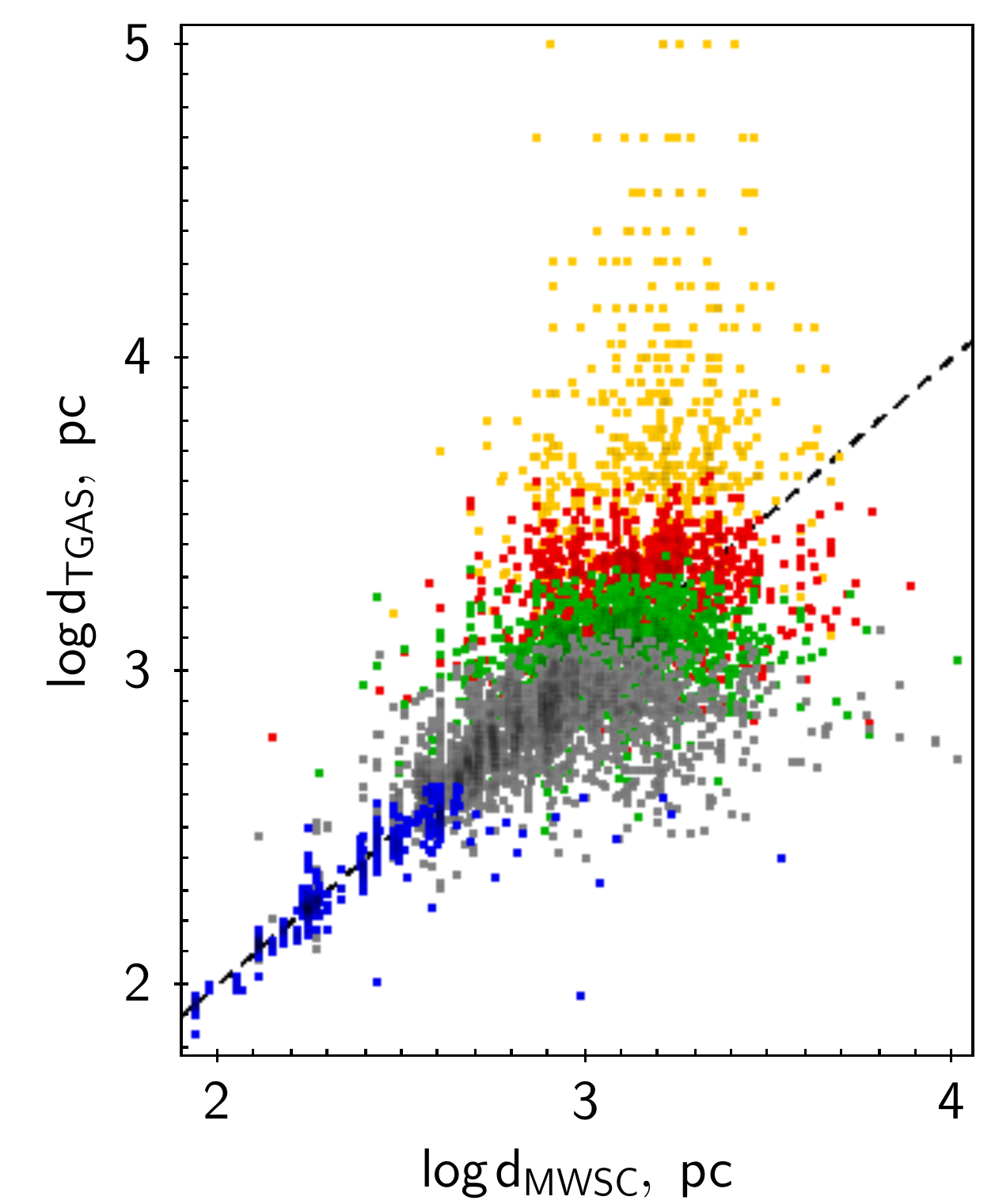}
      \caption{Comparison of photometric (MWSC) and trigonometric (TGAS)
               distances of probable cluster members: yellow
               dots show stars
               with $\varepsilon_{\varpi}/\varpi \ge 1$;
               red dots $0.5 \le \varepsilon_{\varpi}/\varpi < 1$;
               green dots  $0.3 \le \varepsilon_{\varpi}/\varpi < 0.5$;
               grey dots $0.1 \le \varepsilon_{\varpi}/\varpi < 0.3$,
               and blue dots $\varepsilon_{\varpi}/\varpi < 0.1$.
               Stars with $\varpi_{TGAS} \leq 0$ are avoided.}\label{fig:DD}
   \end{figure}

 \begin{figure}
   \centering
   \includegraphics[width=8cm]{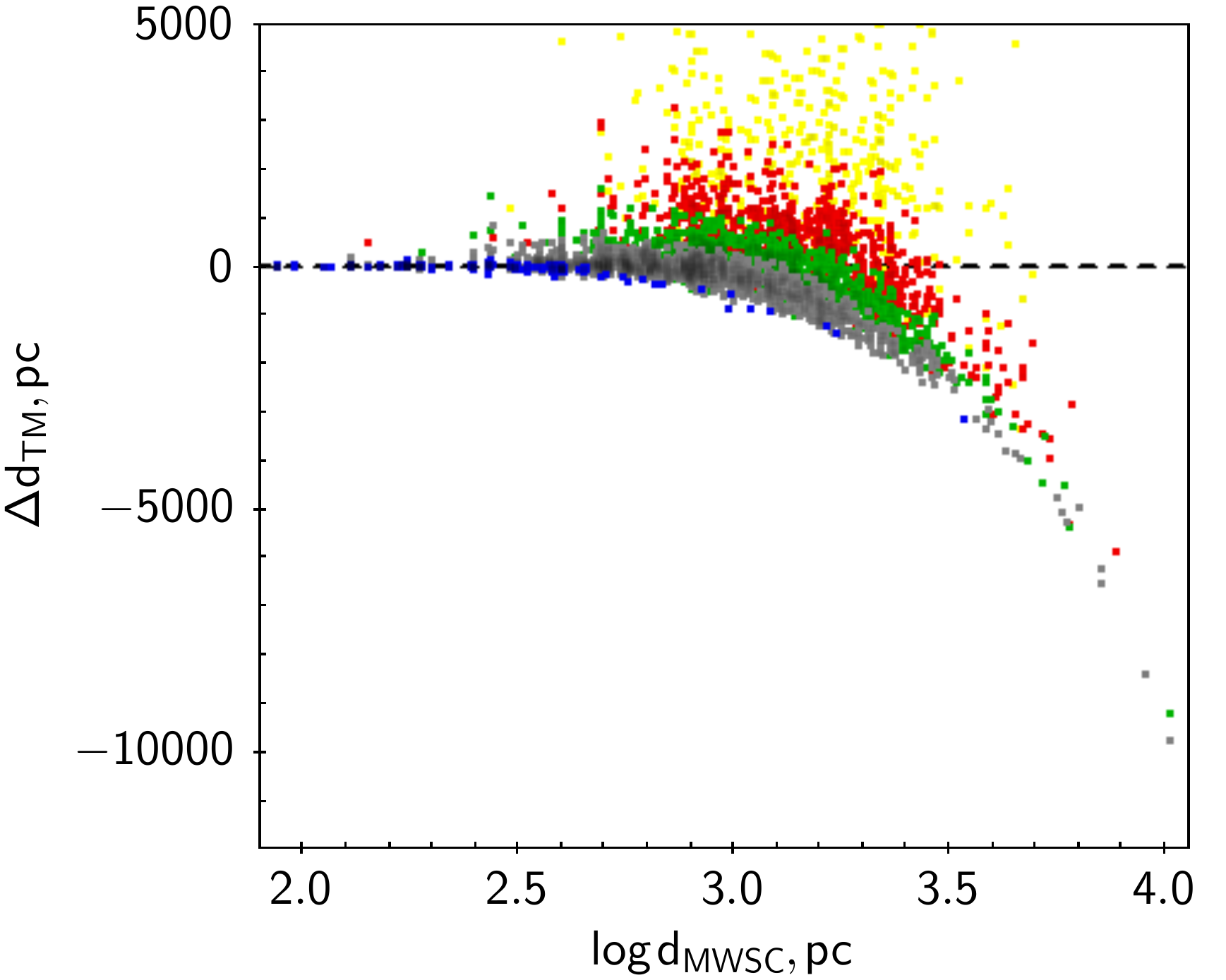}
      \caption{Relation between $\Delta d=d_{TGAS}-d_{MWSC}$
               and $\log d_{MWSC}$ of probable
               cluster members. Colours of the dots are as at Fig.~\ref{fig:DD}.}
      \label{fig:ddeld}
   \end{figure}

\section{Trigonometric versus photometric distances. Comparison, biases, discussion.}
\label{sec:CDB}

In Fig.~\ref{fig:DD} we compare photometric and trigonometric distances of probable cluster members. It is instructive to see, how the stars with different trigonometric parallax accuracy are placed in this figure. In general, both kinds of distances show good agreement until $\log d_{MWSC}\approx 2.8$. At higher photometric distances the spread increases, and a strong asymmetry appears, related to low-quality parallaxes (yellow symbols in the plot). At $\log d_{MWSC}\gtrsim 3$ one can observe another asymmetric bump, below the one-to-one relation. This bump consists of high-quality parallaxes and does not change when we consider subsets with different limits on the relative parallax error. With smaller relative errors of the trigonometric parallaxes the distance range naturally shortens. However, the fraction of higher quality parallax stars declining from agreement at large distances is larger.  This speaks against the hypothesis of a Lutz-Kelker bias as the main reason for the bump, because this statistical effect critically depends on the relative parallax error becoming smaller with better quality parallaxes (see, for the case of a flat distribution of stars, \cite{2017AstBu..72..122R}). We considered these stars as foreground interlopers with correct $d_{TGAS}$, but wrong membership probabilities (and hence their photometric distances). However, we noted that the fraction of such interlopers was small. We did also not see a significant difference in the scatter of the different groups of stars in Fig.~\ref{fig:DD}, when we used the definite cluster members (not shown) instead of the probable cluster members.

To estimate the degree of agreement between the photometric and trigonometric distances, one can use the relation between $\Delta d =d_{TGAS}-d_{MWSC}$ and $\log d_{MWSC}$ (Fig.~\ref{fig:ddeld}). The plot clearly confirms the impression that we already had from Fig.~\ref{fig:DD}. One can see that for $\log d_{MWSC}\lesssim 2.8$ the residuals are small and unbiased indicating very nice agreement of the two scales. At $\Delta d<-1$~kpc there is a long tail of negative residuals $\Delta d$, discussed above as possible foreground interlopers. These stars could be co-moving relatively bright foreground red giants or just affected by uncertain proper motions leading to wrong kinematic MWSC membership probabilities. In contrast, the low-quality parallax dots clearly reside above zero indicating overestimated trigonometric distances due to a low-accuracy parallax bias.

   \begin{figure}
   \centering
   \includegraphics[width=8cm]{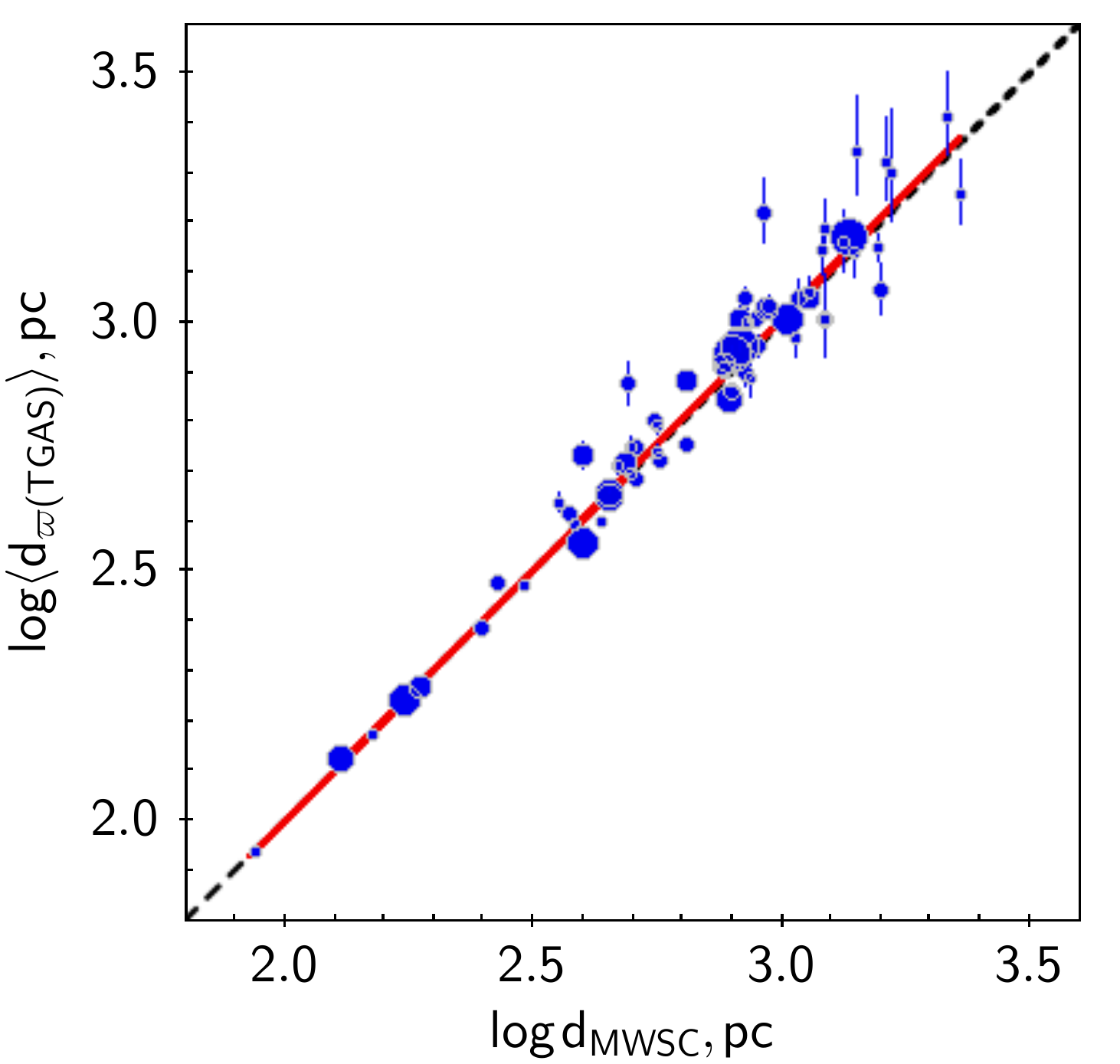}
      \caption{Logarithm of average trigonometric distance of a cluster $\log \langle d_{TGAS}\rangle$ obtained from individual parallaxes of cluster members, vs logarithm of photometric distances $\log d_{MWSC}$. While averaging trigonometric distances, MWSC members out of $ 3 \sigma$ from the mean TGAS distance were discarded from further consideration. 65 clusters with $>$16 members for parallax averaging after this procedure, are shown. The size of the circles is proportional to the number of TGAS parallaxes used for the average calculation; vertical bars represent the errors of the mean TGAS distance. The black dashed line is a bisector, the red line represents the linear regression of the relation.}
         \label{fig:ddClu}
   \end{figure}

Obviously, it should be also instructive to estimate the agreement between trigonometric and photometric distances for the clusters instead of their individual member stars. According to \cite{2017A&A...599A..50A}, \cite{1999ASPC..167...13A}, one should not average distances obtained from inverting observed parallaxes, but instead first average the parallaxes and then invert the result. Therefore, we first determined average TGAS parallaxes and their standard deviations for the members of every populated cluster. Then, we cleaned the datasets from values out of 3 standard deviations (suggesting them to be caused either by foreground co-moving stars or by low quality parallaxes, as discussed above). Our final TGAS cluster distances correspond to the average parallaxes of these cleaned datasets. Fig.~\ref{fig:ddClu} represents the comparison of these distances with their photometric counterparts for the 65 most populated clusters containing 17 and more probable members with TGAS parallaxes, after the removal of outliers. The agreement of trigonometric and  photometric cluster distances is much better than that for individual members and extends much further, up to 1.5\dots2 kpc, if we take into account only populated clusters. The bias is small, and the relation can be appoximated by the weighted linear regression:

\begin{displaymath}
\log \langle d_{TGAS} \rangle = a + b \log d_{MWSC},
\end{displaymath}

\ni where $a = -0.0245 \pm 0.0003$, $b = 1.0102 \pm 0.0001$.

\section{Summary and conclusions}
\label{sec:concl}

We reported the preliminary procedure and results of a direct comparison of the photometric distance scale derived within the MWSC survey with trigonometric distances available for probable members of MWSC clusters in the Gaia DR1 TGAS catalogue. In total, 5743 probable members of 1118 clusters were identified in the TGAS, with 1058 of them having the highest membership probabilities and definitely belonging to 489 clusters, according to the MWSC. We found that the trigonometric and photometric distances of cluster stars perfectly agree up to about 700 pc from the Sun. Beyond this limit, we observed a slowly increasing number of deviating trigonometric or photometric distances. Nevertheless, a good general agreement in both distance scales was seen up to 1.5\dots2 kpc from the Sun, if we used distances of whole clusters with a large number of member stars. Since there were less overestimations of the trigonometric distances, if we considered only the reliable parallaxes, we think that some differences in the distance scales are caused by the uncertainty of trigonometric parallaxes. In case of clearly overestimated photometric distances we attributed the deviation to wrong MWSC membership probabilities, mainly due to uncertain proper motions.

Note that the results of a further improved and final comparison
of the MWSC and TGAS distance scales are published elsewhere
\cite{2017AApL}. In their analysis of the same input data, Kovaleva
et al. \cite{2017AApL} make the comparison between TGAS and MWSC
using parallaxes rather than distances. Their work involves
photometric parallaxes defined as $\varpi_{MWSC}=1000/d_{MWSC}$,
where $\varpi_{MWSC}$ is in mas. The advantage is that these can be
directly compared with all TGAS parallaxes, including negative values,
to avoid any bias. The improved and extended study by Kovaleva
et al. \cite{2017AApL} includes (1) an investigation of TGAS proper
motions of MWSC cluster members, also used in the exclusion of obvious
non-members, (2) a comparison of the MWSC and TGAS parallaxes of
open clusters with the independent results from the Gaia investigation
\cite{2017A&A...601A..19G} of nearby open clusters, (3) a closer look at
some clusters with problematic MWSC parallaxes, and (4) an investigation
of possible small systematic effects in the TGAS-MWSC parallax differences of
open clusters as a function of Galactic longitude.

\bigskip

\ni {\bf Acknowledgements:} This work was partly supported by the Russian Federation of Basic Researches project number 16-52-12027. DK acknowledges support of the Scientific School 9951.2016.2, and thanks Dr. Alexey Rastorguev for valuable comments.

\end{document}